\documentclass[
aps,%
11pt,%
final,%
notitlepage,%
oneside,%
twocolumn,%
nobibnotes,%
nofootinbib,%
superscriptaddress,%
noshowpacs,%
centertags]%
{revtex4}
\usepackage{multirow}
\usepackage{array}
\newcolumntype{L}[1]{>{\raggedright\let\newline\\\arraybackslash\hspace{0pt}}p{#1}}
\newcolumntype{C}[1]{>{\centering\let\newline\\\arraybackslash\hspace{0pt}}p{#1}}
\newcolumntype{R}[1]{>{\raggedleft\let\newline\\\arraybackslash\hspace{0pt}}p{#1}}

\def\squareforqed{\hbox{\rlap{$\sqcap$}$\sqcup$}}

\def\sq{\ifmmode\squareforqed\else{\unskip\nobreak\hfil
\penalty50\hskip1em\null\nobreak\hfil\squareforqed
\parfillskip=0pt\finalhyphendemerits=0\endgraf}\fi}

\def\degr{\hbox{$^\circ$}}

\def\arcmin{\hbox{$^\prime$}}

\def\arcsec{\hbox{$^{\prime\prime}$}}

\def\utw{\smash{\rlap{\lower5pt\hbox{$\sim$}}}}

\def\udtw{\smash{\rlap{\lower6pt\hbox{$\approx$}}}}

\def\fm{\hbox{$.\!\!^{\rm m}$}}

\def\fdg{\hbox{$.\!\!^\circ$}}

\def\farcm{\hbox{$.\mkern-4mu^\prime$}}

\def\farcs{\hbox{$.\!\!^{\prime\prime}$}}

\def\diameter{{\ifmmode\mathchoice
{\ooalign{\hfil\hbox{$\displaystyle/$}\hfil\crcr
{\hbox{$\displaystyle\mathchar"20D$}}}}
{\ooalign{\hfil\hbox{$\textstyle/$}\hfil\crcr
{\hbox{$\textstyle\mathchar"20D$}}}}
{\ooalign{\hfil\hbox{$\scriptstyle/$}\hfil\crcr
{\hbox{$\scriptstyle\mathchar"20D$}}}}
{\ooalign{\hfil\hbox{$\scriptscriptstyle/$}\hfil\crcr
{\hbox{$\scriptscriptstyle\mathchar"20D$}}}}
\else{\ooalign{\hfil/\hfil\crcr\mathhexbox20D}}%
\fi}}

\newcommand{\ab}{Astrophysical Bulletin}

\newcommand{\mnras}{Monthly Notices Roy. Astronom. Soc.}

\newcommand{\pasp}{Publ. Astronom. Soc. Pacific}

\begin{document}
\selectlanguage{english}

\title{ASTRONIRCAM --- INFRARED CAMERA-SPECTROGRAPH \\ for STERNBERG INSTITUTE 2.5 METER TELESCOPE}

\author{\firstname{A.~E.}~\surname{Nadjip}}
\affiliation{Moscow Lomonosov State University, Sternberg Astronomical Institute, Moscow 119234, RUSSIA}

\author{\firstname{A.~M.}~\surname{Tatarnikov}}
\affiliation{Moscow Lomonosov State University, Sternberg Astronomical Institute, Moscow 119234, RUSSIA}

\author{\firstname{D.~W.}~\surname{Toomey}}
\affiliation{Mauna Kea Infrared, LLC, 21 Pookela St., Hilo, HI 96720, USA}

\author{\firstname{N.~I.}~\surname{Shatsky}}
\email{kolja@sai.msu.ru}
\affiliation{Moscow Lomonosov State University, Sternberg Astronomical Institute, Moscow 119234, RUSSIA}

\author{\firstname{A.~M.}~\surname{Cherepashchuk}}
\affiliation{Moscow Lomonosov State University, Sternberg Astronomical Institute, Moscow 119234, RUSSIA}

\author{\firstname{S.~A.}~\surname{Lamzin}}
\affiliation{Moscow Lomonosov State University, Sternberg Astronomical Institute, Moscow 119234, RUSSIA}

\author{\firstname{A.~A.}~\surname{Belinski}}
\affiliation{Moscow Lomonosov State University, Sternberg Astronomical Institute, Moscow 119234, RUSSIA}

\date{\today}
\newcommand{\mkm}{$\mu$m}

\begin{abstract}
ASTRONIRCAM is a cryogenic-cooled slit spectrograph for the spectral range 1--2.5 mkm installed at the Nasmyth focus of the 2.5-meter telescope of the Caucasian observatory of Sternberg Astronomical Institute of Lomonosov Moscow State University. The instrument is equipped with the HAWAII-2RG 2048$\times$2048 HgCdTe array. Grisms are used as dispersive elements. In the photometric mode ASTRONIRCAM allows for extended astronomical object imaging in the field of view of 4.6$\times$4.6 arc minutes with the 0.269~arcsec/pixel scale in standard photometric bands J, H, K and Ks as well as in narrow-band filters CH$_4$, [Fe~II], H$_2$ v=1-0 S(1), Br$_\gamma$ and CO. In the spectroscopic mode, ASTRONIRCAM takes spectra of extended or point-like sources with spectral resolution $R=\lambda/\Delta \lambda \le 1200$. The general design, optical system, detector electronics and readout, amplification and digitization scheme are considered. The conversion factor GAIN measurement results are described as well as its dependence on the accumulated signal (non-linearity).

The full transmission of the atmosphere-to-detector train ranges from 40 to 50\% in the wide-band photometry mode. The ASTRONIRCAM sensitivity at the 2.5-m telescope is characterized by the limiting J=20, K=19 star magnitudes measured with the 10\% precision and 15 minutes integration at the 1~arcsec atmospheric seeing conditions. The references to first results published on the base of ASTRONIRCAM observations are given.

\end{abstract}

\maketitle

\section{Introduction}

ASTRONIRCAM (The ASTROnomical Near InfraRed CAMera) is a camera-spectrograph for the 1.0 to 2.5~microns spectral range which was designed and manufactured by the Mauna Kea Infrared, LLC company\footnote{http://www.mkir.com} in a contract with Moscow Lomonosov State University (MSU) for the new 2.5-meter telescope of the Caucasian Mountain Observatory (CMO) of Sternberg Astronomical Institute (SAI).

CMO SAI is the scientific and educational facility of Moscow State University developed in 2009--2015. The observatory is located at the north-eastern ridge of Mt.~Shatdzhatmaz (25~km to the south of the Kislovodsk resort city, Karachai-Cherkessian Republic; N43\degr~44\arcmin, E42\degr~40\arcmin, 2112~m a.s.l.). Besides the good accessibility and closeness of the city infrastructure, this site is characterized by fairly good astroclimatic conditions. The monitoring held in 2007--2015 at the mountain \cite{kornilov2014,kornilov2016} resulted in the astronomical observations useful average night time quantity of 1320 hours (45\% of the full night time). Out of this amount, nearly 50\% of clear sky time is capable of precision photometric measurements. Atmospheric transparency and its stability at CMO are well characteristic for the sites located at elevations around 2000 meters. Ground temperature and relative humidity measurements deliver the median water vapour precipitate height of 7.7~mm (PWV, \cite{voziakova2012}). The median image quality (seeing) of 0\farcs96 (seconds of arc) at the 0.5\mkm\ wavelength. Being converted by the Kolmogorov law consequent relation $\theta(\lambda) = \theta(\lambda_0) (\lambda/\lambda_0)^{-0.2}$ into infrared domain, this turbulence strength corresponds to 0\farcs8 and 0\farcs7 seeing at wavelengths of 1.1\mkm\ and 2.3\mkm, respectively. The best observational conditions are encountered in the autumn-winter season.

The 2.5-m (F/8) reflector is the main instrument of CMO SAI. The telescope optical system is Ritchey-Chr\'etien design, ASTRONIRCAM is installed in one of four Nasmyth focal stations of the telescope which has an alt-az mount type.

ASTRONIRCAM is designed for taking the direct images and low to medium resolution spectra ($\lambda/\Delta \lambda \le 1200$) of astronomical objects with angular sizes up to 4\farcm6 in standard photometric bands Y, J, H, K and Ks. Apart from wide-band standard filters, a set of narrow-band filters is installed in ASTRONIRCAM for imaging in a number of astrophysically important infrared (IR) lines of $CH_4$ ($\lambda$=1.65\mkm), [Fe~II] ($\lambda$=1.64\mkm), H$_2$ (1-0) S(1) ($\lambda$=2.12\mkm), Br$_\gamma$ and CO ($\lambda$=2.29\mkm). Two linear polarization filters with perpendicular orientation allow for polarimetric measurements.

First light of ASTRONIRCAM happened in May 2015, shortly after the 2.5-m telescope was put in the trial exploitation. During last two years a number of test and scientific observational sets were conducted. This article describes the optical and mechanical instrument design, its detector and electronic read-out system. A number of laboratory and observational measurements of ASTRONIRCAM characteristics are outlined.

\section{Opto-mechanical system}
\subsection{General layout}

ASTRONIRCAM is a cryogenic-cooled instrument\footnote{With exception of the detector unit and a few other details, ASTRONIRCAM is a twin instrument of the TIRSPEC camera-spectrograph being in operation at the 2-m Himalayan Chandra Telescope (HCT), Hanle (Ladakh), India, see http://www.tifr.res.in/$\sim$daa/tirspec/ .} built according to the classical long-slit IR spectrograph scheme and equipped with the large format array detector HAWAII-2RG 2048$\times$2048 HgCdTe ($\lambda\!<\!2.6$\mkm). The basic components of the optical system are the three-lens collimator, four-lens camera objective and three automated turrets which contain the spectral slits (and one field diaphragm), photometric filters and dispersers (represented by grisms). The use of grisms which disperse light along the system optical axis instead of reflective diffraction gratings traditionally used  in the infrared allowed to unite the camera and spectrograph in a single instrument without substantial optical system and design complication. Switching between photometry and spectroscopic modes and back is made by insertion of respective grisms and/or filters in the parallel beam.

In order to reduce the instrumental background from the intrinsic thermal radiation of optical and mechanical elements, the whole optical system is placed inside the vacuum cryostat and cooled down together with photometric filters and the detector to the working temperature Т$\approx$77--80~K.

\begin{figure*}
\setcaptionmargin{0mm}
\onelinecaptionstrue
\includegraphics[width = 12cm, keepaspectratio]{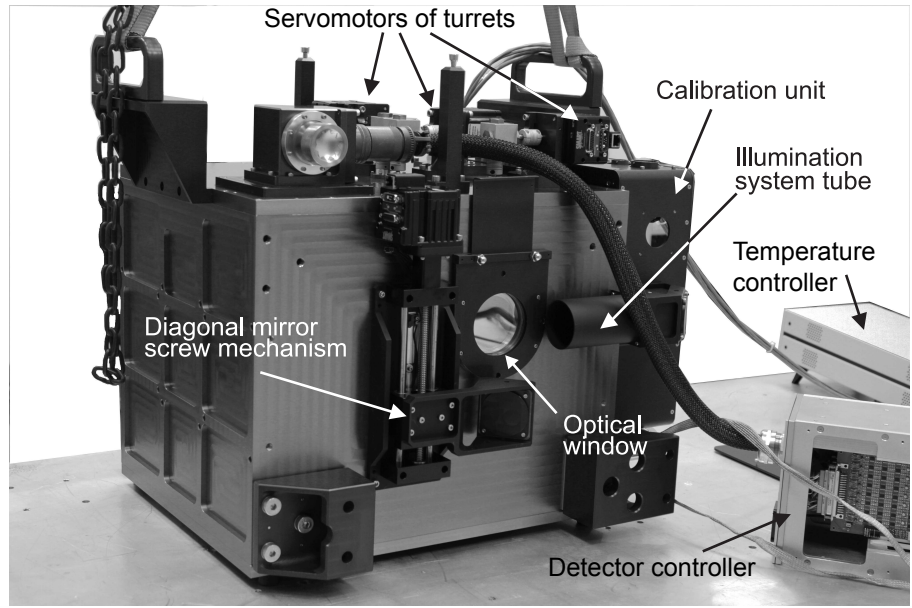} 
\captionstyle{normal}
\caption{External view of ASTRONIRCAM camera-spectrograph\label{fig:anc1}}
\end{figure*}

In Fig.~\ref{fig:anc1} the external view of the ASTRONIRCAM system is shown. The basic block of the instrument is the liquid nitrogen cooled rectangular-shaped cryostat. The optical window of the cryostat is located in the front wall of the body. To the right, the tube of the calibration unit illumination system is seen in the figure. Inside the tube, the achromatic condenser is housed which projects the light source onto the spectral slit inside the cryostat. The light beam from the tube is reflected towards the cryostat with help of a diagonal mirror which is translated in front of the window with a motor-driven screw mechanism.

\begin{figure}
\setcaptionmargin{0mm}
\onelinecaptionstrue
\includegraphics[width = 8cm, keepaspectratio]{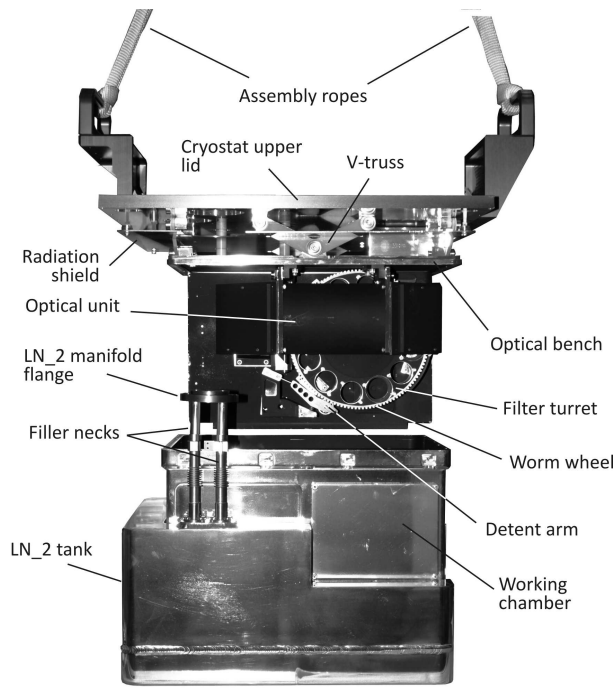}
\captionstyle{normal}
\caption{Installation of the optical system into the cryostat working chamber\label{fig:anc2}}
\end{figure}

Some important units of the internal composition are shown in Fig.~\ref{fig:anc2}. The modular optical system is assembled on the rigid optical bench which is attached to the cryostat upper lid via stiff V-shaped trusses from G-10 fiberglass (having high temperature stability and low thermal conductivity). The figure shows the moment of installation of the optical system inside the working chamber of the cryostat manufactured together with the liquid nitrogen tank. The optical bench closes the working chamber from the top and after screwing has a good thermal contact with the nitrogen so the elements installed on the bench are effectively cooled down to about 80--82~K. Filling the tank with liquid nitrogen (LN$_2$) is accomplished via a couple of filler necks with a manifold that is hermetically connected to the tank via thin stainless steel goffered tubes.

The detector is installed in a separate unit which is thermally insulated from the rest of the optical unit and cooled down to 78--80~K (depending on the LN$_2$ tank filling degree) via a separate pure copper strap. The strap is attached to the detector base (manufactured from molybdenum which has a low thermal expansion coefficient and high thermal conductivity) at one side and to the nitrogen tank lid at another side. Thermal insulation is made with help of V-trusses from G-10 which attach the detector unit to the previous optical unit with the camera lens. This means of attachment and cooling of the detector installed on a lightweight and low thermal capacity base provides the possibility of precise automatic thermal stabilization\footnote{Currently the active temperature stabilization is not used.}. For the regulation and measurement of the detector temperature the thermal sensor and heating resistor are installed at the molybdenum base while the temperature control is performed with external Lakeshore 335 controller. Another thermal sensor is installed at the optical bench to monitor the temperature of spectrograph optical elements. 

In order to minimize the heat inflow to the detector, its leads are wired to the 61-contact hermetic connector installed at the top of the cryostat via two flexible flat cables having thin conductors from manganine which has extremely low thermal conductivity and low thermo-emf. It is worth noting that in spite of the low manganine thermo-emf (0.1 to 1.0$\mu$V/1K when coupled to copper) the contact EMF may turn out significant due to large temperature difference between the detector leads and connector (around 200~K) and may also vary depending on currents in operation which in turn influences the detector voltage levels.

Outside the cryostat, at the top of the vacuum jacket, three servomotors with embedded controllers\footnote{http://www.animatics.com, model SM2316D-ETH} are attached which are geared to the focal and two filter turrets. Hermetic input of rotation from the drives to the turrets inside the vacuum jacket is made via the ferro-liquid high vacuum feed-throughs\footnote{http://www.ferrotec.com, model SS-188-SLES}. The axles rotate the turrets via the screw drives which consist of the molybdenum plastic self-lubricating worm and the tooth rim formed directly on the turret wheel. Setting of the turret in a working position is performed with help of a spring-actuated detent mechanism equipped with the Hall-effect sensors to control the position. The fixing element of the mechanism is a bearing on a springed arm which drops in triangle grooves of dedicated positions. The original design of the worm unit which allows for free axial movement of the worm between two limits on the axle makes the wheel precisely positioned by the bearing in a groove when any tangential force from the worm is absent.

In Table~\ref{tab:cryostat_param} some basic cryostat parameters are listed. The dimensions are given both for the vacuum jacket alone and for the fully equipped system at the focal station where the length (L) is measured along the optical beam entering the instrument.

\begin{table}
\setcaptionmargin{0mm}
\onelinecaptionstrue
\captionstyle{normal}
\caption{Technical characteristics of the cryostat\label{tab:cryostat_param}}
\medskip
\begin{tabular}{|R{5.4cm}|L{2.2cm}|}
\hline
Vacuum jacket size, W$\times$L$\times$H, mm	& 483$\times$400$\times$400 \\
System dimension, W$\times$L$\times$H, mm 		& 625$\times$500$\times$560 \\
Cold mass (without LN$_2$)	& 32 kg \\
Weight of on-telescope assembly (w/o LN$_2$)&	110 kg \\
LN$_2$ tank volume		&	8.6 l \\
LN$_2$ consumption for first cool-down & $\approx$ 40 l \\
Detector cool-down rate	& $<\!1$K/minute \\
Thermal stabilization time @T=78~K (first cool-down)	& 16 hr \\
Working temperature			& 78..80 K \\
LN$_2$ refill periodicity	& 48 hr \\
\hline
\end{tabular}
\end{table}

\subsection{Optical system}
\begin{figure*}
\setcaptionmargin{0mm}
\onelinecaptionstrue
\includegraphics[width = 12cm, keepaspectratio]{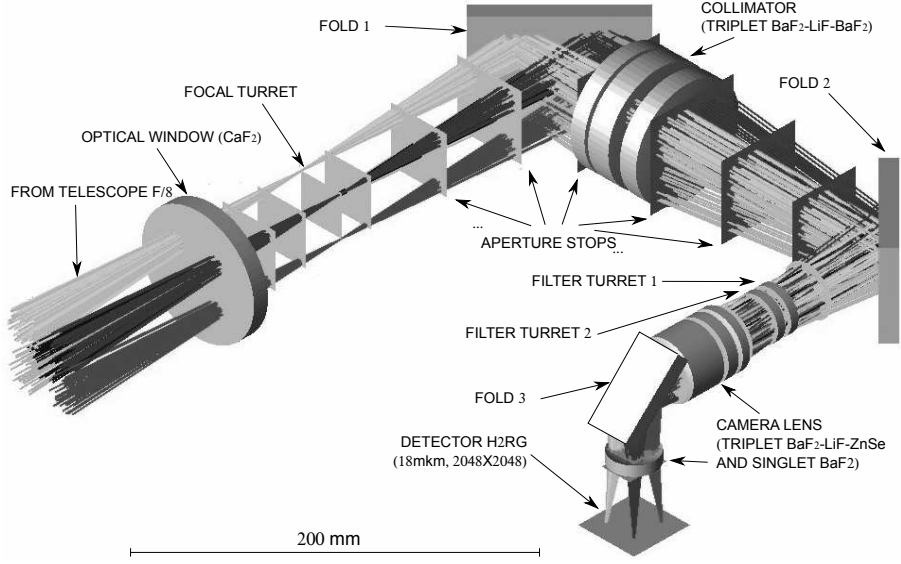}
\captionstyle{normal}
\caption{ASTRONIRCAM optical layout\label{fig:anc3}}
\end{figure*}

In Fig.~\ref{fig:anc3} the general optical layout of the ASTRONIRCAM system is shown. The telescope focal plane is located inside the cryostat. The F/8 converging beam from the 2.5-m telescope enters the optical window (CaF$_2$) and builds an image of observed astronomical object in the focal plane scaled as 10\farcs31/mm. The focal turret contains 11 positions of which 10 are changeable spectral slits and one is a square field diaphragm of the 27~mm side used for the direct imaging mode. The angular size of the field of view is 4\farcm64$\times$4\farcm64 in this mode.

The BaF$_2$-LiF-BaF$_2$ collimator lens triplet ($F_{col} \approx 207$~mm) located one focal length behind the telescope focal plane collimates the light and forms a reimaged telescope pupil of 26.0~mm diameter.

Two filter turrets, each having 12 positions, are located near the intermediate pupil plane. The first (upper) turret hosts wide band J, H, K and Ks filters, two cross-dispersing grisms (YJxd and HKxd, mounted together with their order sorter filters YJsort and HKsort) and two additional order-sorting filters, Ysort and Jsort. The second (lower) turret contains the main YJHK disperser grism, a number of narrow-band filters and two polarizers with mutually perpendicular polarization planes. Both filter turrets have one position open; the second one also contains the cold blank-off used for taking dark current frames and avoiding detector casual saturation.

The filters all have the 4~mm thickness and effective (light) diameters of 26.5~mm. In order to avoid multiple reflections between filter surfaces and subsequent optical elements, the filters are installed with a tilt angle of 5\degr. Due to limited filter apertures, some vignetting (up to 3--4\%) is encountered for marginal rays near the corners of the field of view which decreases slightly the detector illumination.

The four-lens camera objective (triplet BaF$_2$-LiF-ZnSe + singlet BaF$_2$; $F_{cam} \approx 144$~mm) refocuses the parallel beams onto the science detector surface making the observed object image with a scale of 14\farcs87/mm. With such a reduction factor, the array pixel step of 18.5\mkm\ (see below) converts effectively into 0\farcs275/pixel on the sky. This scale, obtained from the optical system computer modeling, slightly differs from  the astrometry calibrated plate scale, 0\farcs269/pixel.

This scale is well adequate for the atmospheric conditions at CMO in the working wavelength domain since the turbulent star image disks are covered by 2--3 pixels depending on the band and actual conditions.

Across the full field of view, the optical system exhibits nearly diffraction quality. The distortion accounts for not more than 1\% at the image margins.

Three flat mirrors are introduced in the optical system for folding the path in a more compact configuration to accommodate it in a relatively small cryostat. The reflecting surfaces of folding mirrors (manufactured from the fused silica) are coated with gold, the material which has an excellent IR reflectivity and durability.

Optical surfaces of the entrance window and all optical lenses are wide-band anti-reflection coated. In order to inhibit scattered light inside the optical train  (due to remaining reflexes and scattering at the optical surfaces) 16 blocking baffles are installed along the optical system path.

\subsection{Filters}

For the purposed of wide band photometry ASTRONIRCAM is equipped with J, H, K and Ks filters of the MKO-NIR system recipe. This photometric system \cite{tokunaga2003}, developed for ground-based infrared observations by a joint effort of Mauna Kea, Gemini and other observatories\footnote{See also http://www.ifa.hawaii.edu/$\sim$tokunaga/ MKO-NIR\_filter\_set.html}, provides the best photometric precision and linear dependence of counts on the airmass compared to other filter sets, both for observations at high-altitude sites (such as Mauna Kea, 4200~m a.s.l., median humidity of 2~mm PWV) and for moderate altitude observatories like CMO (2100 m a.s.l., PWV from 4 to 8~mm in autumn to spring period).

The Table~\ref{tab:filters} contains the characteristics of all ASTRONIRCAM filters manufactured by the Materion company\footnote{https://materion.com/products/precision-optics/precision-optical-filters/, USA}: the central wavelength CWL, the full profile width at the half-maximal transmission level FWHM, the average transmission in the band computed as $T_{avg}=\int P(\lambda)d\lambda$/FWHM and the maximal transmission T$_{max}$. The transmission curves\footnote{http://lnfm1.sai.msu.ru/kgo/instruments/filters/} were recorded by the manufacturer at 77~K and normal incidence, so the wavelengths in the table are corrected to account for 5\degr\ incidence on the filters (maximal shift is $-3$~nm for the K band).

\begin{table}
\setcaptionmargin{0mm}
\onelinecaptionstrue
\captionstyle{normal}
\caption{Filters characteristics\label{tab:filters}}
\medskip
\begin{tabular}{|l|cccc|}
\hline
Filter	& CWL 	& FWHM 	& $T_{avg}$	& $T_{max}$  \\
		&  [nm]	&  [nm]	& [\%]  	& [\%] \\
\hline
\multicolumn{5}{|c|}{ Wide-band photometry filters} \\
\hline
J	& 	1249	&	166	&	88	&	91	\\
H	&	1635	&	291	&	96	&	97	\\
Ks	&	2143	&	303	&	89	&	92	\\
K	&	2191	&	316	&	92	&	94	\\
\hline
\multicolumn{5}{|c|}{ Narrow-band photometry filters}	\\
\hline
CH$_4$ Off &	1581	&	57.2	&	85.5	&	83	\\
$\left[\mathrm{Fe II}\right]$	&	1642	&	26.1	&	97.6	&	96	\\
CH$_4$ On &	1651	&	64.7	&	99.2	&	96	\\
H$_2$ v=1-0 (S1) & 2129	&	46.2	&	94.7	&	92	\\
Br$_\gamma$ &	2165	&	21.2	&	93.4	&	90	\\
Kcont	&	2270	&	39.3	&	91.4	&	90	\\
CO	&	2282	&	30.2	&	93.1	&	90	\\
\hline
\multicolumn{5}{|c|}{ Order sorter filters}	\\
\hline
Ysort	&	1109	&	178	&	83	&	88	\\
Jsort	&	1342	&	271	&	94	&	96	\\
YJsort	&	1263	&	467	&	87	&	95	\\
HKsort	&	1984	&	957	&	91	&	95	\\
\hline
\end{tabular}
\end{table}

\subsection{Grisms}

The Table~\ref{tab:grisms} provides the parameters of grisms installed in ASTRONIRCAM and ruled\footnote{Bach company, USA, http://www.bachresearch.com} directly on the zinc selenide prism back surface (refractive index n=2.4). 
\begin{table}
\setcaptionmargin{0mm}
\onelinecaptionstrue
\captionstyle{normal}
\caption{Grisms parameters\label{tab:grisms}}
\medskip
\begin{tabular}{|c|C{1.2cm}|C{1.2cm}|C{2.5cm}|}
\hline
Grism	& Wedge angle, [\degr]	& grooves density, [mm$^{-1}$] at~80~K	& Blaze wavelength [\mkm] in working orders $m$\\
\hline
YJHK	&	 21.94	&	81.0 	& 6.6  m=1 \\
 &&&2.2  m=3 \\
 &&&1.65 m=4 \\
 &&&1.32  m=5 \\
 &&&1.10  m=6 \\
HKxd	&	 5.00	&	65	& 2.0  m=1 \\
YJxd	&	 8.00	&	162.4	& 1.25  m=1 \\
\hline
\end{tabular}
\end{table}

The YJHK grism is used for taking single order spectra in bands Y, H, H and K in the single-dispersion mode. All the bands are covered by the same grism with help of dispersion in different orders: third, fourth, fifth and sixth for central (blaze) wavelengths of 2.20\mkm, 1.65\mkm, 1.32\mkm\ and 1.10\mkm, respectively. The 3-rd and 4-th order spectra are isolated from adjacent orders contamination by standard K and H filters, while 5-th and 6-th order spectra are rectified by filters Jsort and Ysort, respectively (see Table~\ref{tab:orders}). These non-standard order-sorter filters are used due to the fact that 5-th and 6-th order spectra have marginal wavelengths slightly different from the standard Y and J bands.

The grisms HKxd and YJxd (cross-grisms) are used together with the YJHK grism for the cross-dispersed spectral mode. This mode allows for taking two bands spectra simultaneously: H and K or Y and J. For rectification of HK and YJ spectra the non-standard wide band filters HKsort and YJsort are used (installed in the same cells as the cross-grisms HKxd and YJxd themselves). Table~\ref{tab:orders} shows the spectral ranges for different spectral observation modes.

\begin{table*}
\setcaptionmargin{0mm}
\onelinecaptionstrue
\captionstyle{normal}
\caption{Spectral modes and coverage\label{tab:orders}}
\medskip
\begin{tabular}{|c|c|C{2.5cm}|c|C{2.5cm}|}
\hline
Grism	& Filter	& Spectral order	& CWL [\mkm] 	& Spectral range, [\mkm] \\
\hline
YJHK	& Ysort	 & 6		& 1.12	& 1.02--1.2 \\
YJHK	& Jsort	 & 5		& 1.33	& 1.21--1.48 \\
YJHK	& H	 & 4		& 1.65	& 1.49--1.78 \\
YJHK	& K	 & 3		& 2.2	& 2.04--2.35 \\
YJHK+YJ	& YJsort & 6, 5 xd 1	& 1.25	& 0.95--1.50 \\
YJHK+HK	& HKsort & 4, 3 xd 1	& 2.0	& 1.46--2.50 \\
\hline
\end{tabular}
\end{table*}

\subsection{Spectral slits. Spectral resolving power}

The focal turret contains five long spectral slits with angular length of 275\arcsec\ and angular widths of 0\farcs9, 1\farcs3, 1\farcs8, 2\farcs7 and 7\farcs2 and five short slitlets with the angular length of 9\arcsec\ and same widths. Short slits are designed for the cross-dispersed spectral mode.

The slit with the angular width 1\farcs3 matches well the atmospheric conditions at CMO in the near-IR range. At the wavelengths 1.25\mkm\ and 2.2\mkm\ the median seeing is 0\farcs8 and 0\farcs7, respectively. In such conditions, the 1\farcs3 slit lets more than 94\% of the point source energy through at $\lambda\!=\!1.25$\mkm\ and more than 97\% of energy at $\lambda\!=\!2.2$\mkm. The thinnest slit of 0\farcs9 may be efficiently used at the best seeing conditions while wider slits of 1\farcs8, 2\farcs7 and 7\farcs2 may appear helpful for low surface intensity extended objects or for spectrophotometry.

It is easy to show that for a slit instrumental profile the spectral resolving power of a grism spectrometer coupled to a telescope may be computed as follows (see, e.g. \cite{beckers2003}):
\begin{equation}
R \equiv \frac{\lambda}{\Delta\lambda}\approx(n - 1) \frac{d_{gr}}{D_{tel}\theta_s} \tg\beta, \label{eq:1}
\end{equation}
where $n$ is the grism refractive index  ($\approx 2.4$), $d_{gr}$ is the grism effective aperture (26.5~mm), $D_{tel}$ is the telescope aperture diameter (2500~mm), $\theta_s$ is the slit angular size (projected onto the sky, in radians) and $\beta$ is the grism wedge angle (21\fdg94). Calculated by this formula, the values of the spectral resolving power are following: $R_{0.9} \approx 1370$, $R_{1.3} \approx 920$, $R_{1.8} \approx 690$, $R_{2.7} \approx 460$, $R_{7.2} \approx 170$ for slits of respective widths. Meanwhile the equation \ref{eq:1} does not account for diffraction and aberration effects which cause widening of the slit image and may lead to substantial increase of effective slit widths and respective decrease of resolving power. Computer modeling of our optical system demonstrated that diffraction and aberrations (significantly amplified with grisms introduced in the beam) lead to sensible widening of slit images only for the most narrow slit (0\farcs9). As result, the maximal resolving power drops down to $R_{0.9} \approx 1200$ for central wavelengths (close to blaze condition) in all spectral bands which is confirmed by direct measurement of the calibration emission lamp spectrum. Since aberrations grow towards the order edges, effective slit width slightly grows as well towards marginal wavelengths and towards ends of a long slit, slightly differently for various wavelengths and focusing.

For wider slits the diffraction and aberrations cause mainly some dithering of the slit edge images without significant increase of its effective width.

\subsection{Spectral calibration unit}

The illumination system of the spectral calibration unit contains an integrating sphere fed by two light sources --- the Argon lamp with an emission spectrum and the tungsten lamp as a continuous light source. An opal glass plate 

The spectral slits accept illumination from the source via the two-lens condenser projecting the integrating sphere output port onto the focal plane with help of a gold coated diagonal mirror (see Fig.\ref{fig:anc1}).

\subsection{Optical system transmission}
\subsubsection{Photometric mode}
In Table~\ref{tab:transmission} estimates of the telescope and instrument transmission in the photometric mode are given. Values were derived from the manufacturer data on the telescope and instrument folding mirrors reflectivity, filters transmission and losses at instrument optical system component AR coatings which are made on the optical window and lenses. The telescope aperture obstruction by the secondary baffle and spider arms are taken into account which bring an effective linear obstruction ratio of 0.41.

\begin{table}
\setcaptionmargin{0mm}
\onelinecaptionstrue
\captionstyle{normal}
\caption{Optical system transmission coefficients\label{tab:transmission}}
\medskip
\begin{tabular}{|c|c|c|c|c|c|}
\hline
\multirow{2}{*}{Band}	& \multirow{2}{*}{Telescope}	& \multicolumn{2}{c|}{Camera}	& \multicolumn{2}{c|}{Full path} \\
\cline{3-6}
& & Calc. & Meas. & Calc. & Meas. \\
\hline
J	& 	0.69	&	0.58	&	0.59	&	0.40	&	0.41	\\
H	&	0.73	&	0.63 	&	0.70	&	0.46	&	0.51	\\
K	&	0.75	&	0.59 	&	0.64	&	0.44	&	0.48	\\
\hline
\end{tabular}
\end{table}

The last column of Table~\ref{tab:transmission} presents estimates of the full system transmission which were obtained by measurement of MKO photometric system standard star fluxes. Observations were performed in clear nights by stars at small zenith distances. Since the optical elements and atmospheric transmission is not so well defined, the agreement of measured and predicted quantities is fairly good. These estimates may well be used for exposure calculation for stars of different magnitudes.

\subsubsection{Spectral mode}
Typical grism transmission at the blaze wavelengths is $T_{max}\approx0.8$ and usually declines to the margins of the free dispersion range by some 50\%. If we take $T_{gr}\approx0.6$ as an average grism transmission in the free spectral range, we derive thus the best case estimates of the full system transmission in the singe and cross-dispersed spectral modes as 0.22/0.13, 0.26/0.16 and 0.25/0.15 for the J, H and K bands, respectively. These numbers do not take into account scattering at uncoated facets of a diffraction grating. Preliminary measurement results show the fraction of the light scattered off the grooves as high as some 30\% but this is even not the full measure of losses at grisms in use.

Currently there is still not enough observational data to derive firmly the real spectral mode optical transmission of ASTRONIRCAM.

\section{Detector}
\label{sec:detector}
As it was already mentioned above, ASTRONIRCAM is equipped with the HAWAII-2RG 2048$\times$2048 format detector with the HgCdTe photodiodes array and the cutoff wavelength of about 2.6\mkm\ manufactured by the Teledyne Scientific and Imaging company. This is one of the best contemporary large format infrared arrays widely used in last decade both for ground-based observations as well in space-born near-infrared experiments. At the working temperature of 77~K the detector H2RG has a nearly 100\% quantum efficiency, extremely low rate of dark current (about 0.02 electrons per second) and low read-out noise (around 12 electrons for a single or correlated double sampling read-out, see below).

The H2RG detector has a redundant format with respect of the ASTRONIRCAM optical system designed for the use with smaller $1024\times1024$ arrays. That is why only the central part of H2RG is effectively used in our instrument. Nevertheless, we made this choice for ASTRONIRCAM because of extremely good H2RG characteristics and also the potential to use nearly all vertical detector format in the spectral mode (along the dispersion).

Photo-sensitive elements of the H2RG detector array are the photo-diodes on PN junction (p-on-n). The array has square 18$\times$18~\mkm\ pixels which are spaced at the grid with a step of 18.5 \mkm\ in both coordinates of a square array. 

The detector is manufactured using the traditional hybridization technology for infrared arrays when a photo-sensitive HgCdTe plate with a grid of photo-diodes is glued over the surface of the silicon CMOS multiplexor (MUX) integrated scheme \cite{loose2003}. The diodes are electrically coupled to their MUX elements via indium bumps. The key to success in reaching the highest quantum efficiency of the detector is the removal of the CdZnTe substrate upon the hybridization at which the crystalline HgCdTe plate was grown. The detector-ascendants had this substrate serving as an optical window through which the array was exposed; this window  absorbed a significant amount of light and caused interference pattern (``etaloning'') in images.

\begin{figure}
\setcaptionmargin{0mm}
\onelinecaptionstrue
\includegraphics[width = 8cm, keepaspectratio]{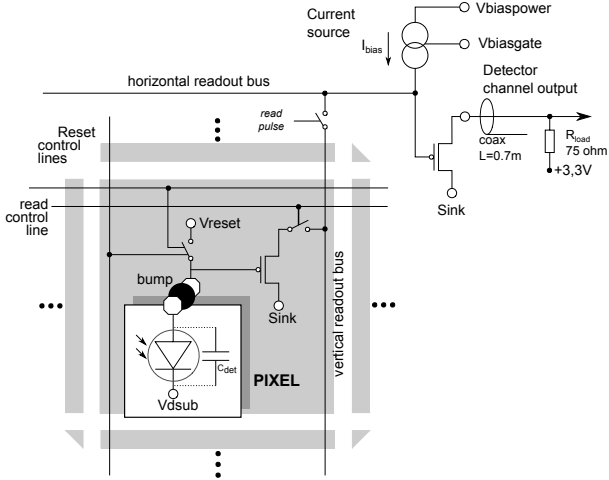}
\captionstyle{normal}
\caption{Detector pixel read-out simplified scheme\label{fig:SFD}}
\end{figure}

In Fig.~\ref{fig:SFD} the photo-diode schematics is shown which represents the Source Follower per Detector (SFD) approach or direct photo-electrons integration scheme \cite{fossum1993} (the charges produced in the photo-diode are integrated directly on its intrinsic capacitance C$_{det}$).

Photo-diodes work with a certain biasing (in the diode mode) but without a permanent connection to the external power source. The biasing of the photo-diode is created by feeding the short voltage pulse Reset to the lead connected to the gate of a MOS-transistor of the SF, with an amplitude +V$_{rst}$ being less than the permanently maintained +V$_{dsub}$ voltage at the base.  As result, the photo-diode receives the inverse biasing which value is defined by the difference of the reset and base voltages: $\mathrm{V}_{bias} = \mathrm{V}_{rst} - \mathrm{V}_{dsub}$.

The value of V$_{rst}$ (actually the potential, since the electrical circuit is open) at the signal lead of the photo-diode defines the zero level of the detector signal. When the cell illumination takes place, the physical processes inside the p-n junction produce the positive growth of voltage at the signal lead proportionally to the intensity and exposure time of illumination. Respectively, together with the signal growth, the value of the initial p-n junction biasing effectively decreases towards zero and further to small positive values (forward bias) when the voltage at the signal lead ceases to grow (saturation voltage). Thus, the initial value of the photo-diode biasing determines the working range of the detector signal. We work with the following settings: V$_{dsub}\!=\!0.6$~V, V$_{rst}\!=\!0.25$~V and hence the initial biasing is V$_{bias}\!=\!-0.35$~V.

Since the detector has neither optical nor electrical shutter, in order to maintain the needed zero signal level in the idle mode, the Reset pulses are applied periodically.

Source followers do not have individual load resistors but are fed by the common current source to which the MUX connects their sources at the moments of signal read-out. The current source stabilizes the current of the transistor source during read-out and removes non-linear skews of the video-pulses formed. The video-pulses from the outputs of source followers are transmitted to the detector output via the buffered amplifier of current and further via the coaxial cable at the input of the video-processor board of the detector controller. The buffered current amplifier is used to match the output impedance of photo-diode SFs with the 75~ohm coaxial cable impedance (loaded to the 75~ohm resistor). This efficiently cancels the cable influence on the video-pulses formation process at outputs of SFs.

The MUX performs the detector array read-out sequentially pixel by pixel along the array row and row by row in vertical (slow) direction with a sampling rate of 300 thousand pixels per second. In the same manner and rate the array reset is performed at the exposure beginning. The identical timing of resetting and read-out of the array guarantees the equal exposure duration for all the array elements.

While reading the array through a single channel with the rate 300 kHz all the elements (2048$\times$2048 pixels) are read in 14.6 seconds. Higher frame rate is attained by division of the array at several vertical stripes and reading them simultaneously using independent read-out channels. The H2RG MUX has 32 channels while we use only 4 of them by which four vertical areas sized 512$\times$2048 pixels are digitized simultaneously. This choice requires a significantly simpler circuitry and provides a fairly low thermal generation of the detector. In such a way, the whole array readout time reduces to 3.646~s which defines the minimal exposure time. Meanwhile, the MUX provides not only the full-frame readout scheme, but also the windowing mode. While having a smaller window size, one may reach even shorter exposures\footnote{Windowing mode is performed using only one channel though.}.

Photo-diode source followers have very high input resistance which results in effectively zero leakage of the accumulated charge through the MOS-transistor. That is why during readout of the signal from outputs of SFs the accumulated photo electrons remain all in the cell. This provides the possibility to perform the non-destructive readout (NDR) of the detector array during exposure. This capability is exploited in well known methods of infrared image acquisition such as Correlated Double Sampling (CDS) or Ramp Sampling \cite{fowler1991, robberto2007}. The first method (CDS) involves taking two frames, at the beginning and at the end of exposure, and plain subtraction of the latter from the former. Both frames are burdened by a spatial kTC noise pattern which is generated during the array resetting before exposure and retains intact during the exposure. Thus, subtraction of frames efficiently removes this pattern from the final image where the signal difference represents integral signal in each pixel.

The second method ``Ramp'' implies periodic array read-out during the whole exposure duration. For each pixel one obtains the sample points equally spaced in time domain which allow determination by the least squares method the regression coefficient of the signal growth. The regression allows to derive the growth rate, the total accumulated signal and even non-linearity factors. This method allows to cancel the kTC-noise in the output data like in the CDS approach but also to diminish the effective read-out noise and filter out reading glitches, cosmics etc. We use both methods in observations. Table~\ref{tab:h2rg} summarizes the important parameters of our H2RG detector.

\begin{table}
\setcaptionmargin{0mm}
\onelinecaptionstrue
\captionstyle{normal}
\caption{Parameters of ASTRONIRCAM H2RG detector\label{tab:h2rg}}
\medskip
\begin{tabular}{|R{5.6cm}|L{2cm}|}
\hline
Detector type & p-on-n photo-diode \\
\hline
Sensitive material &  HgCdTe \\
\hline
Cut-off wavelength (by 50\% of maximum level) & 2.59 \mkm \\
\hline
Working temperature & 77 K \\
\hline
Full format dimension & 2048$\times$2048 \\
\hline
Effectively used format & 1024$\times$1024 \\
\hline
Pixel size & 18$\times$18\mkm \\
\hline
Pixel step, both dimensions & 18.5\mkm \\
\hline
Quantum efficiency (QE) in 0.6--1.0 \mkm\ range & 78\% \\
\hline
Quantum efficiency (QE) in 1.0--2.4 \mkm\ range & 94\% \\
\hline
Median dark current rate at V$_{bias}\!=\!0.25$~V and 77~K& 0.02 e/s \\
\hline
Median read-out noise (CDS, 300~kHz) & 12 e \\
\hline
Pixel electric capacitance at V$_{bias}\!=\!0.25$~V & 40 fF \\
\hline
Pixel well depth at V$_{bias}\!=\!0.25$~V & 120700 e \\
\hline
Median electron-to-voltage conversion factor & 3.7 $\mu$V/e \\
\hline
\end{tabular}
\end{table}

\section{Controller of detector}

Detector is operated by the ARC Gen III controller manufactured by Astronomical Research Cameras, Inc\footnote{San Diego, USA, http://www.astro-cam.com}. This controller is also widely known to astronomical community as ``Leach controller'' (by the author name, Dr. Robert Leach, \cite{leach2000}) or ``SDSU controller'' (San Diego State University).

The controller provides the detector with all the required power voltages, generates all the control and clocking pulses which are fed to the detector MUX, and further performs the amplification and filtering the video-signals as well as their digitization, scrambling and forming the data packages for the control computer.

Amplification and analog-to-digit conversion of the video-signals which are read-out from four detector channels and conducted at the input of the controller by four coaxial cables is performed by four parallel signal processors which are part of the controller video-board ARC-46. Each video-processor is a multi-stage DC current amplifier with an analog-digital converter (ADC) at the output. The input stage of the processor is an instrumental amplifier with the unity amplification coefficient on the base of three operational amplifiers (OA). This precision buffered amplifier has high input resistance, low intrinsic noise and high degree of dumping of common-mode interference. Its output is fed to the inverted input of the subsequent OA with the amplification coefficient  K$_{oa}\!=\!5$. The non-inverting OA input is connected to the offset voltage U$_{off}\!=\!-2$~V in order to subtract the DC component of the video-pulses which they have due to offset of output voltages of photo-diode source followers and of the buffered MUX amplifier. At the OA output the zero level is manifested at approximately middle of the useful signal range so when the detector signal grows the pulses polarity is inverted at some signal level.

The next amplification stage is the OA-based integrator which is equipped with the Sample and Hold keys. The sampling is made synchronously with the video-pulses at the 300~kHz rate. The sampling duration defines the video-pulse integration time while the duration of holding the result is limited by the time required by the subsequent ADC to digitize the signal. Besides the role of a low-frequency filter, this electronic unit bears another important function to cut the ``parasitic'' pulses which are generated at the buffered MUX amplifier at the moments of keys switching. This is made by a small delay of sampling with respect to the video-pulse front.

The amplification coefficient of the integrator equals the ratio of the integration duration to its time constant. Integrator has a switch allowing to select one of two time constants, 1~$\mu$s or 4~$\mu$s. We work with 1~$\mu$s time constant and integration duration of $t=2\mu$s, so the integrator voltage amplification factor is K$_{int} = t/\tau = 2/1 = 2$.

Since all the subsequent cascades of the video-processor including ADC have a unity amplification factor, the full voltage amplification coefficient of the video-processor is K$_{pr}$ = K$_{oa} \times$ K$_{int} = 5 \times 2 = 10$.

Output integrator voltages (protected by limiter diodes) are within the range of $-2.5$~V to $+2.5$~V. The voltage $+2.08$~V corresponds to the offset zero level of the detector signal while $-2.5$~V is a maximal registered signal level of the detector. Thus, while the detector signal grows, the voltage at the video-processor amplifier output is not growing but decreasing (from $+2.08$~V to $-2.5$~V). 

The analog-to-digital conversion is performed by the 16-bit differential ADC. The special ADC driver scheme is used to match the non-symmetric (``single-wire'') output of the integrator with the differential input of the ADC. This driver converts the single-polarity (positive or negative) voltages into symmetric (counter-phased bipolar) differential voltages. The input differential voltages range of ADC is from $-2.5$ to $+2.5$~V. Input of $-2.5$~V corresponds to the minimal digital result (zero Analog-to-Digital units, ADU) while $+2.5$~V is converted into the maximal resulting number of 65535~ADU ($2^{16} -1$). Correspondingly, the ADC resolution by input voltage is 5 / 65535 = 76.3~$\mu$V/ADU. This value, being backward converted to the  video-processor input equals to 76.3 [$\mu$V/ADU] / K$_{pr}$ = 7.63 [$\mu$V/ADU].

The digitized data from four outputs of video-processors (4 ADCs) enter the buffered electronic FIFO unit which forms the digital frames which are further transmitted via the optical link interface to the control computer.

\section{Conversion factor}

Digital images contain data in relative digit units, ADU. The relation between the digital data and physically measured quantities (numbers of detected photo-charges in detector cells) is manifested by the conversion factor of the electronic system which is usually designated as GAIN. The GAIN value is defined as a ratio of accumulated photo-charges to the corresponding digital amount: GAIN$=N_{det}[e] / N_{adc}$[ADU] and thus has the dimension [e/ADU]. This is actually the inverse conversion factor if we recall the terminology commonly accepted in electronics.

The photo-charges to digits conversion process may be divided in two stages. At the first stage, the charges generated by the photo-diode are converted into the voltage at the intrinsic cell capacitance with a conversion coefficient K$_{pd}$ = q$_e$/C$_{p-n}$~[$\mu$V/e] where q$_e$ is electron charge and $C_{p-n}$ is the barrier capacitance of the p-n photo-diode junction ($C_{det}$ in Fig.~\ref{fig:SFD}). At the second stage this voltage is transmitted via the source follower of the photo-diode, buffered amplifier of the MUX and coaxial cable to the video-processor input, then amplified by the latter and converted into the digital form with the total conversion factor of the electronic system К$_{es}$ [$\mu$V/ADU]. The GAIN coefficient is expressed via these two coefficients as GAIN = K$_{es}$ /K$_{pd}$. 

\begin{figure}
\setcaptionmargin{0mm}
\onelinecaptionstrue
\includegraphics[width = 8cm, angle=0, keepaspectratio]{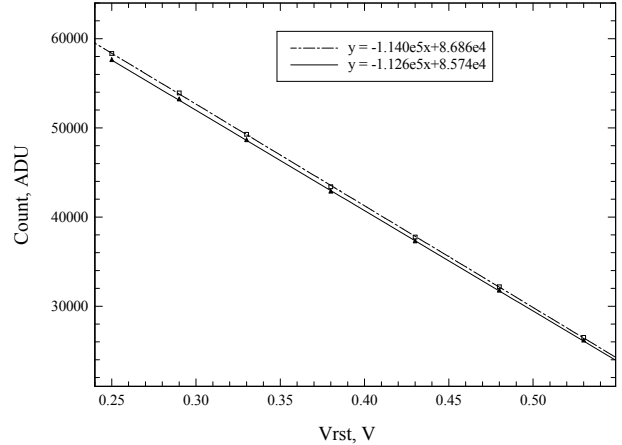}
\captionstyle{normal}
\caption{Digital output relation on the voltage (potential) value at the signal leads of the photodiodes of 2-nd (squares) and 3-rd (triangles) channels\label{fig:vreset}}
\end{figure}

The value of K$_{es}$ may be measured directly applying different values of V$_{rst}$ voltage to the signal leads of the detector photo-diodes and plotting the resulting digital outputs after respective short dark exposures against V$_{rst}$. The reciprocal proportionality coefficient gives directly the K$_{es}$ estimate. In Fig.~\ref{fig:vreset} the respective plots for the second and third array read-out channels are given consisting of values averaged over the 500$\times$1000 pixel areas of these channel stripes (which are in the central exposed zone of the detector). Both plots demonstrate a fairly good linearity of the electronic system for both channels. It is also seen that two channels have slightly different zero biases which correspond to the working mode with V$_{rst}$\,=\,0.25~V.

The plots show also different reciprocal slopes which envisages the respective difference in transfer coefficients of these channels: K$_{es}$(2) = 8.77 [$\mu$V/ADU] and K$_{es}$(3) = 8.88 [$\mu$V/ADU] for the second and third channels, respectively. This un-equality is most likely due to transfer coefficients of the MUX buffered amplifiers of these channels. The value of the resulting transfer coefficient of the MUX and signal coaxial cable to the video-processor may be calculated as a ratio of the ADC resolution as taken at the input of the video-processor (7.63 $\mu$V/ADU) to the derived full transfer coefficients K$_{es}$(2) and K$_{es}$(3): k(2) = 7.63/8.77 = 0.870 and k(3) = 7.63/8.88 = 0.860 for the second and third channels, respectively. In particular, this implies that the voltage amplification coefficients of these channels are K(2) = k(2)$\times$K$_{pr}$ = 0.87$\times$10 = 8.70 and K(3) = k(3)$\times$K$_{pr}$ = 0.86$\times$10 = 8.60. 

As stated above, besides the measured values of K$_{es}$(2) and K$_{es}$(3), the conversion factor of the photo-diode K$_{pd}$ is required as well. The manufacturer reports the capacitance $C_{p-n}$ of array photo-diodes equal to $40\!\times\!10^{-15}$~farad at the inverse bias voltage 0.3~V. Accepting this value, we obtain K$_{pd}$=4~[$\mu$V/e] and, consequently, GAIN(2) = K$_{es}$(2)/K$_{pd}$ = 8.77 [$\mu$V/ADU] / 4 [$\mu$V/e] = 2.19 [e/ADU] and GAIN(3) = K$_{es}$(3)/K$_{pd}$ = 8.88 [$\mu$V/ADU] / 4 [$\mu$V/e] = 2.22 [e/ADU] for the channel 2 and 3, respectively.

The Table~\ref{tab:gains} summarizes the values of the electronic system amplification and conversion coefficients discussed above.

\begin{table*}
\setcaptionmargin{0mm}
\onelinecaptionstrue
\captionstyle{normal}
\caption{Amplification and conversion coefficients of electronic system\label{tab:gains}}
\medskip
\begin{tabular}{|r|C{1.2cm}|C{1.2cm}|}
\hline
Channel number & 2 & 3 \\
\hline
Video-processor voltage amplification coefficient	&	10	&	10	\\
Multiplexor and cable conversion coefficient	&	0.870	& 	0.860	\\
Full amplification coefficient of electronic system		&	8.70	&	8.60	\\
Electronic system conversion coefficient [$\mu$V/ADU]&	8.77	& 	8.88	\\
Intrinsic conversion coefficient of photo-diodes [$\mu$V/e]	& 4	&	4	\\
GAIN [e/ADU]	&	2.19	&	2.22	\\
\hline
\end{tabular}
\end{table*}

The value of the GAIN coefficient is one of the most important characteristic of the electronic system since many other relevant parameters like the overall instrument efficiency are derived through this value. That is why we performed a number of additional electronic system tests in order to obtain independent GAIN estimates using the commonly used statistical method of measurement the relation between the signal variance and the average counts number. The essence of this method is simple. It is easy to show that while converting the counts of the signal accumulated during a certain exposure time in detector cells into numbers (Poisson statistics) the counts variance $D$[ADU] and the mathematical expectation of the average count $M$[ADU] are related as following: $D = \mathrm{k}M$, where k is a conversion coefficient. Recall that GAIN is a reciprocal conversion coefficient, so GAIN$=1/$k. Hence, given the experimentally determined slope of this D--M relation, one may obtain the GAIN coefficient, with some assumptions on the signal non-linearity and spacio-temporal test source stability. 

We plotted these graphs of the variance against average expressed in [ADU] using two data retrieval approaches: with the spatial sampling across the array pixels and with the temporal sampling for each particular pixel with multiple identical detector exposure series.

In experiments aimed at determination of GAIN it is important to maintain the stable level of light intensity falling onto detector which directly affects both the average level and the measured variance. As a source we took the metal plate of the room temperature put in front of the cryostat optical window and took all the measurements in the Br$_\gamma$ filter. The instrument resided in the isolated lab room with a stable temperature. The degree of constancy of the illumination was assessed by the measurement of the signal itself which was averaged across the working area of the detector. In our experiments, the root mean square variance of the averaged signal through the experimental series constituted $\sim0.1$\% of the signal level so we may easily neglect the signal variance in our conclusions.

\subsection{Temporal sampling}
In order to obtain the representative statistics we performed 50 cycles of the detector illumination of equal duration. In each cycle, the array was read-out by the Ramp-method with 14 NDR frames (see section~\ref{sec:detector}), each frame having its own level of integrated signal. The persistence effect inherent for many infrared arrays (after-glowing of previous exposures signal in images obtained later) was taken into account by making a 10~minutes pause between exposures when the light was blocked by a cold blank and the RESET pulses were cleaning the array periodically. The static kTC noise pattern was removed by subtraction of the first NDR from all subsequent frames (like in CDS method; this noise is about 23--25~ADU in our detector). Since the variance is caused both by the quantum photo-charges noise and read-out noise (D = D$_{qua}$ + D$_{ro}$), the NDR frames with the average signal level above $\approx$5000~ADU were selected so the read-out noise could be neglected (we measured read-out noise in lab conditions to be 12.5--13.5~e). For having the non-linearity effects also limited (see section~\ref{sec:nonlin}) the upper signal was restricted to be $<$17000~ADU.

\begin{figure}
\setcaptionmargin{0mm}
\onelinecaptionstrue
\vspace{3mm}
\includegraphics[width = 8cm, keepaspectratio]{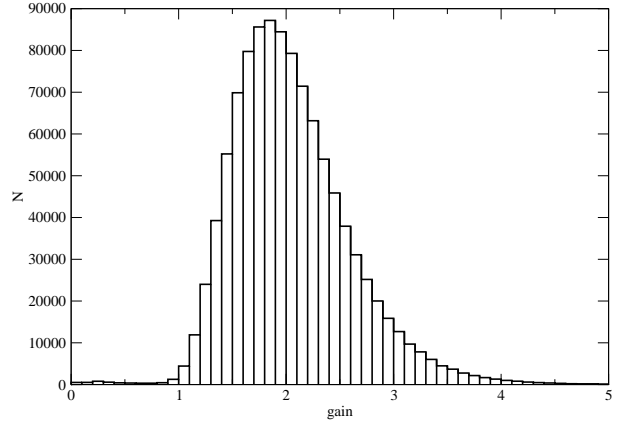}
\captionstyle{normal}
\bigskip
\caption{Histogram of distribution of the $gain$ value in pixels of the working area of the array\label{fig:gain_spatial}}
\end{figure}

A set of 50 NDR frames was taken for each NDR-number (having a certain illumination level) in order to calculate the average value and the variance of counts for each array pixel. Each pixel thus obtained its own D--M relation and the related $gain$ coefficient (lowercase letters denote the quantities related to pixels while area-averaged values are designated by capital letters). In Fig.~\ref{fig:gain_spatial} we present a histogram of the derived gain values for the central array area of 1000$\times$1000 pixels. A similarly shaped distribution was obtained by Ninan et al for the TIRSPEC instrument detector \cite{tirspec}.

The $gain$ distribution shown in the histogram has a very wide width far exceeding the expected gain variance from the real signal distribution in pixels of the detector array area in use where the root mean square relative count deviation is no more than 3\%. High distribution variance is explained by the limited sample data volume. In order to obtain the representative sample capable of revealing the individual pixel gain deviations its volume should be inflated by a factor of several tens which is not feasible technically. This also means that average and median GAIN values from this histogram are statistically biased. Meanwhile, this average gain in the 2-nd and 3-rd channels GAIN$_{avg}$(2,3) = 2.18~e/ADU is close to the numbers obtained with another sampling method described below.

\subsection{Spatial sampling}
To implement the spatial sampling we used the same 50 Ramp-frames as in the temporal sampling. From this data set we composed different pairwise combinations and for each pair of frames with the same integration time we computed the average and variance of pixel-wise count differences across the working areas of a second and third channel separately. Again, like in the previous calculation, we used only the frames with the integrated signal level between 5000 and 17000~ADU. Given the variance and average signal level for each pair of frames, we computed the GAIN value for this pair. Averaged over the set of different exposure pairs, this gave us GAIN$_{avg}$(2) = 2.23$\pm$0.02 and GAIN$_{avg}$(3) = 2.26$\pm$0.03~e/ADU for the second and third channels, respectively.

The values obtained with the spatial sampling technique agree fairly well with GAIN measurements from V$_{rst}$ variation described above. One observes some systematic shift of the values by 0.04 in direction of higher values (1.8\% in relative measure) which may be attributed to a deviation of a real barrier photo-diode capacitance from the tabular value (0.7~fF, respectively).

\section{Non-linearity}
\label{sec:nonlin}
\begin{figure}
\setcaptionmargin{0mm}
\onelinecaptionstrue
\includegraphics[width = 8cm, keepaspectratio]{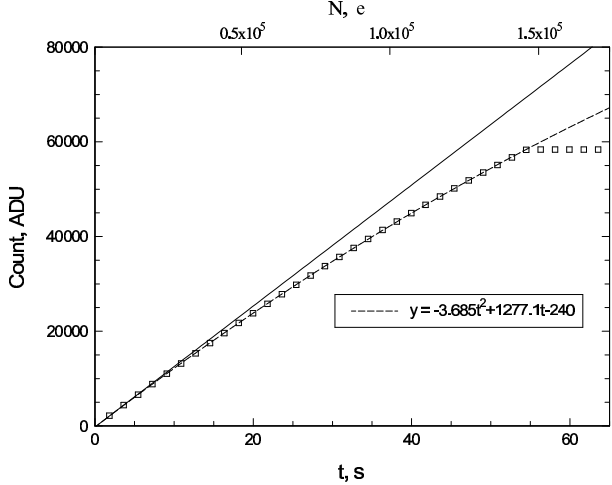}
\captionstyle{normal}
\caption{Signal growth in a sample pixel with time (lower abscissa axis) and accumulated electrons number (upper axis), the approximating relation (dashed line) and vertex line for the lowest signal part (solid line)\label{fig:signal_vs_t}}
\end{figure}

The GAIN values obtained above refer to the limited range of integrated counts (5000--17000 ADU). We deliberately made this restriction keeping in mind the relation of GAIN on the detector signal level. In Fig.~\ref{fig:signal_vs_t} the dependence of the accumulated signal count on the exposure duration is presented (averaged over many experimental series) taken at constant illumination level. Thus we expect the photo-charges generation rate also constant so this curve is a manifestation of non-linearity of the photo-charges to digits conversion in the whole range of signal levels (covering some 58000~ADU). The flat part of the graph at high signals corresponds to saturation of the ADC happening when the photo-diodes are approaching their complete discharge (see section~\ref{sec:detector}).

Earlier above it was shown (Fig.~\ref{fig:vreset}) that the amplification and digitization electronic system is highly linear. Thus we come to conclusion that the GAIN non-linearity is due to purely the non-linearity of the photo-charges-to-voltage conversion inside the photo-diodes themselves.

\begin{figure}
\setcaptionmargin{0mm}
\onelinecaptionstrue
\includegraphics[width = 6cm, keepaspectratio]{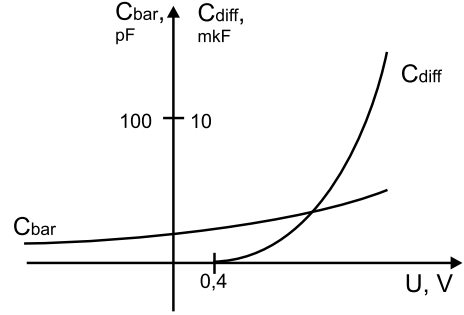}
\captionstyle{normal}
\caption{Example of a capacitance versus voltage characteristic of a p-n junction. Here C$_{bar}$ is barrier capacitance and C$_{dif}$ is diffusion capacitance\label{fig:volt-farad}}
\end{figure}

Such a behavior of the conversion coefficient K$_{pd}$ = q$_e$/C$_{p-n}$ [$\mu$V/e] has a simple explanation. The photo-diode capacitance C$_{p-n}$ is a barrier capacitance of p-n junction which value depends non-linearly on the biasing voltage of the junction. When the bias voltage is decreasing (during accumulation of photo-charges at the cell capacitance and respective increase of the electric potential at the signal lead of the photo-diode) the value of the barrier p-n junction capacitance grows. This dependence is manifested by the capacitance--voltage characteristic of a p-n junction which example is shown in Fig. ~\ref{fig:volt-farad}.

\begin{figure*}
\setcaptionmargin{0mm}
\onelinecaptionstrue
\begin{tabular}{ccc}
\includegraphics[width = 7.8cm, height=5.4cm]{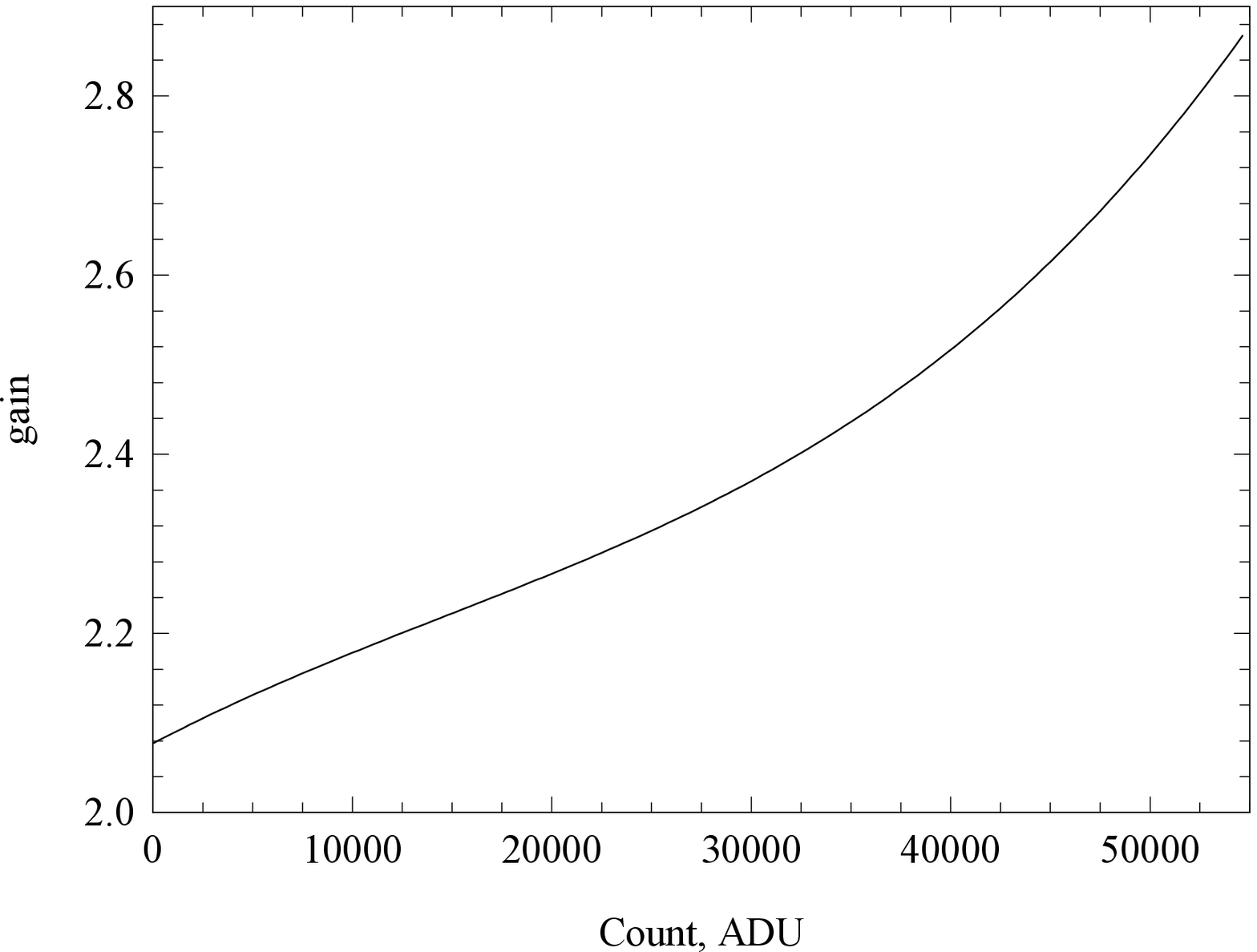} & \hspace{3mm} &
\includegraphics[width = 7.8cm, keepaspectratio]{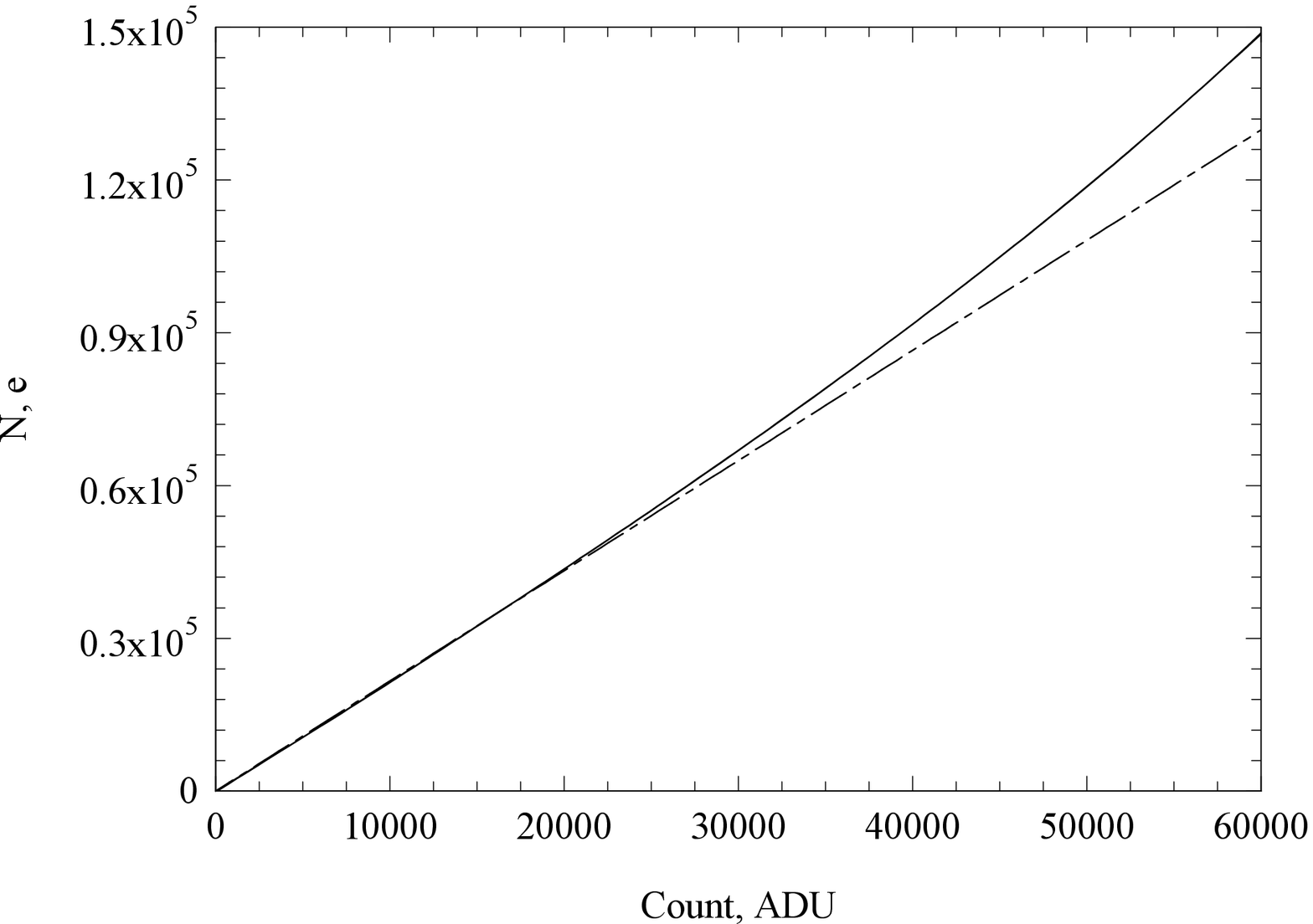} \\
\end{tabular}
\captionstyle{normal}
\caption{Dependence of gain (left) and accumulated number of photo-electrons (right; dash-dotted line depicts linear trend at low signals) on the signal level\label{fig:gain_vs_signal}}
\end{figure*}

In order to determine the dependence of the GAIN coefficient on the signal level using the data given in Fig.\ref{fig:signal_vs_t} we made a variable change turning the time abscissa argument into the number of photo-charges accumulated  at the photo-diode capacitor over the integration time $t$ (upper abscissa axis). In such a graph conversion the value of GAIN for a given signal level is defined as a coefficient of the reciprocal slope of the approximating function (its reciprocal derivative) in a given point. We scaled the photo-charges axis in such a way that at the signal 10500~ADU the GAIN equals exactly 2.18~e/ADU which is the value derived for the 5000-17000~ADU signal range.  This way we derive the GAIN versus signal graph shown in Fig.~\ref{fig:gain_vs_signal} together with its cumulative (integral) representation.

In practice, the non-linearity correction of the detector employs the polynomials of various degree (from 3 \cite{hilbert2008} to 9 \cite{chil2015}). The correction is performed with a relation determined for each channel, row-wise or pixel-wise. In our work we use the pixel-wise correction by 4-th order polynomials.

\section{Sensitivity}
In order to estimate the limiting magnitudes attainable by ASTRONIRCAM with the 2.5-meter telescope of CMO SAI we used the frames taken at low zenith distance in night 6--7 November 2016 with a stable and good transparency (visual extinction $\approx$0.25 and the infrared extinction $\approx$0.05) and good seeing (FWHM $\le1$\arcsec\ in IR). The resulting image was combined from 30 separate Ramp-frames each obtained with exposure time 30 seconds and small (a few arcsec) shifts of the telescope attitude between exposures (the so called \emph{dithering} method). This way the total integration time of the summed frame in each of J, H and K filters constituted 900~sec. In these resulting frames the stars of J=20\fm2, H=19\fm4 and K=18\fm8 were measurable at the signal-to-noise level of about 15. Given other error sources, the photometric accuracy is about 0\fm1.

\section{Colour transformation equations}

By reducing observations of 20 standard stars from the Leggett et al. list \cite{leggett2006} performed in nights with stable transparency we derived photometric equations of the reduction from our instrumental into the standard MKO system. Here below we only quote the result of this work while the detailed description of observation conditions and data reduction shall be published later. The colour transformation equations look like following:
\begin{equation}
\Delta (J-K) = 0.989 \Delta (j-k)
\end{equation}
\begin{equation}
\Delta (J-H) = 0.987 \Delta (j-h)
\end{equation}
\begin{equation}
\Delta (H-K) = 0.996 \Delta (h-k)
\end{equation}
where J, H and K are star magnitudes in MKO while j, h and k are those in the instrumental system.

\section{Conclusions}

The laboratory experiments performed to determine the working characteristics of ASTRONIRCAM and some results of trial observations demonstrated that being operated with the SAI 2.5-meter telescope this instrument holds a valuable scientific potential and may effectively be used for educational programs as well as for fundamental scientific research of different type astronomical objects.

Due to limited volume of this article we deliberately omitted some important instrumental characteristics like the detector persistence, the BIAS frames stability, optical and electrical ``ghosts'' (due to so called ``cross-talk'' between cells or channels) and some others which affect the data photometric precision. Spectroscopic characteristics were also only outlined. We continue to explore these issues and will summarize the results in coming papers.

The first scientific results obtained with ASTRONIRCAM embrace observations of sources of very different nature, both galactic and extra-galactic. Some were already published in \cite{oknyansky2017,tatarnikova2016,berdnikov2017,lamzin2017} as well as in a number of Astronomical Telegrams.

\acknowledgements
Authors are grateful to our colleagues M.~A.~Burlak for help in performing the photometric measurements and their processing, B.~S.~Safonov for discussion and statistical interpretation of the detector gain measurement results as well as the engineer personnel of the Caucasian SAI observatory for continuous support of the works at the telescope. NS, ACh, SL and AB thank the Russian Scientific Foundation for the grant (2) 17-12-01241 which partially supported their work.

Authors thank Federal State Unitary Enterprise ``Foreign Trade Association ``Vneshtechnika'' for diligent execution of the ASTRONIRCAM instrument delivery, purchased from the funds of the Program of Development of the Moscow University.
%
%
{}
%
%
%
%
%

\pagebreak

\end{document}